\documentstyle[12pt,graphicx]{article}
\begin{document}
\baselineskip 0.6cm

\def\simgt{\mathrel{\lower2.5pt\vbox{\lineskip=0pt\baselineskip=0pt
           \hbox{$>$}\hbox{$\sim$}}}}
\def\simlt{\mathrel{\lower2.5pt\vbox{\lineskip=0pt\baselineskip=0pt
           \hbox{$<$}\hbox{$\sim$}}}}

\begin{titlepage}

\begin{flushright}
UCB-PTH-07/25
\end{flushright}

\vskip 2.4cm

\begin{center}

{\Large \bf 
Flavorful Supersymmetry
}

\vskip 1.0cm

{\large Yasunori Nomura$^{a,b}$, Michele Papucci$^c$, and 
Daniel Stolarski$^{a,b}$}

\vskip 0.4cm

$^a$ {\it Department of Physics, University of California,
          Berkeley, CA 94720} \\
$^b$ {\it Theoretical Physics Group, Lawrence Berkeley National Laboratory,
          Berkeley, CA 94720} \\
$^c$ {\it School of Natural Sciences, Institute for Advanced Study,
          Princeton, NJ 08540}

\vskip 1.2cm

\abstract{Weak scale supersymmetry provides elegant solutions to many 
of the problems of the standard model, but it also generically gives 
rise to excessive flavor and $CP$ violation.  We show that if the 
mechanism that suppresses the Yukawa couplings also suppresses flavor 
changing interactions in the supersymmetry breaking parameters, 
essentially all the low energy flavor and $CP$ constraints can be 
satisfied.  The standard assumption of flavor universality in the 
supersymmetry breaking sector is not necessary.  We study signatures 
of this framework at the LHC.  The mass splitting among different 
generations of squarks and sleptons can be much larger than in 
conventional scenarios, and even the mass ordering can be changed. 
We find that there is a plausible scenario in which the NLSP is a 
long-lived right-handed selectron or smuon decaying into the LSP 
gravitino.  This leads to the spectacular signature of monochromatic 
electrons or muons in a stopper detector, providing strong evidence 
for the framework.}

\end{center}
\end{titlepage}

\section{Introduction}
\label{sec:intro}

Despite many new alternatives, weak scale supersymmetry is still regarded 
as the leading candidate for physics beyond the standard model.  It not 
only stabilizes the electroweak scale against potentially large radiative 
corrections, but also leads to successful gauge coupling unification 
at high energies and provides a candidate for dark matter.  The fact 
that supersymmetry must be broken, however, leads to a severe flavor 
and $CP$ problem.  Including generic supersymmetry breaking parameters 
of order the weak scale causes flavor changing and $CP$ violating 
processes with rates much greater than current experimental bounds. 
In fact, the problem has become more severe because of recent 
experimental progress, especially in $B$ physics.

The most common approach to this problem is to assume that supersymmetry 
breaking and its mediation to the supersymmetric standard model sector 
preserve flavor.  In other words, the mechanism for generating the 
Yukawa couplings for the quarks and leptons is separate from that 
which mediates supersymmetry breaking, so the fundamental supersymmetry 
breaking parameters do not contain any sources of flavor or $CP$ 
violation.  This is typically achieved in one of two ways.  The 
first is to simply assume flavor universality and $CP$ conservation 
for the supersymmetry breaking parameters at the scale where the 
low energy field theory arises~\cite{Chamseddine:1982jx}, and the 
second is to impose a low energy mechanism which leads to flavor 
universality~\cite{Dine:1981gu,Dine:1994vc}.

A careful look at the problem, however, shows that the situation does 
not need to be as described above.  We know that the Yukawa couplings 
for the first two generations of quarks and leptons are suppressed, 
implying that there is some mechanism responsible for this suppression. 
Suppose that this mechanism suppresses all non-gauge interactions 
associated with light generation fields, not just the Yukawa couplings. 
Then the supersymmetry breaking masses for the light generation squarks 
and sleptons are also suppressed at the scale where the mechanism is 
operative, suppressing flavor and $CP$ violation associated with these 
masses.  This scenario was considered before in Ref.~\cite{Kitano:2006ws} 
in the context of reducing fine-tuning in electroweak symmetry breaking. 
The necessary flavor universal contribution to the squark and slepton 
masses arises automatically at lower energies from the gaugino masses 
through renormalization group evolution.  An additional contribution 
may also arise from a low energy mechanism leading to flavor universal 
supersymmetry breaking masses.

In this paper we study a scenario in which the physics responsible for 
the quark and lepton masses and mixings is also responsible for the 
structure of the supersymmetry breaking masses.  We call this scenario 
flavorful supersymmetry in order to emphasize the direct connection 
between flavor physics and supersymmetry breaking.  To preserve the 
success of gauge coupling unification in the most straightforward way, 
we assume that this physics lies at or above the unification scale 
$M_U \approx 10^{16}~{\rm GeV}$.  We find that, in contrast to naive 
expectations, a large portion of parameter space is not excluded by 
current experimental data.  We study implications of this scenario 
on the low energy superparticle spectrum, which can be tested at 
future colliders.  In particular, we point out distinct signatures 
at the LHC, arising in the plausible case where the gravitino is 
the lightest supersymmetric particle.  Throughout the paper we assume 
that $R$ parity is conserved, although the framework can be extended 
straightforwardly to the case of $R$ parity violation.

The organization of the paper is as follows.  In section~\ref{sec:framework} 
we describe our basic framework, and in section~\ref{sec:constraint} 
we show that it satisfies experimental constraints from low energy 
flavor and $CP$ violation.  In section~\ref{sec:implication} we discuss 
implications on the weak scale superparticle spectrum, and we analyze 
signatures at the LHC in section~\ref{sec:LHC}.  Finally, conclusions 
are given in section~\ref{sec:concl}.

\section{Framework}
\label{sec:framework}

Suppose that the supersymmetric standard model, or supersymmetric grand 
unified theory, arises at a scale $M_*$ ($\simgt M_U$) as an effective 
field theory of some more fundamental theory, which may or may not be 
a field theory.  We consider that the physics generating the Yukawa 
couplings suppresses all non-gauge interactions associated with the quark, 
lepton and Higgs superfields $Q_i$, $U_i$, $D_i$, $L_i$, $E_i$, $H_u$ 
and $H_d$ (and $N_i$ if we introduce the right-handed neutrinos), where 
$i=1,2,3$ is the generation index.  In particular, it suppresses the 
operators generating the supersymmetry breaking masses at the scale $M_*$:
\begin{eqnarray}
  {\cal L} &=& \Biggl( 
    \sum_{A=1,2,3} \int\! d^2\theta\, \eta_A 
      \frac{X}{M_*} {\cal W}^{A\alpha} {\cal W}^A_\alpha 
    + {\rm h.c.} \Biggr) 
\nonumber\\
  && {} + \int\! d^4\theta \Biggl[ 
    \kappa_{H_u} \frac{X^\dagger X}{M_*^2} H_u^\dagger H_u 
    + \kappa_{H_d} \frac{X^\dagger X}{M_*^2} H_d^\dagger H_d 
\nonumber\\
  && \qquad {} + \Biggl( \kappa_\mu \frac{X^\dagger}{M_*} H_u H_d 
    + \kappa_b \frac{X^\dagger X}{M_*^2} H_u H_d 
    + \eta_{H_u} \frac{X}{M_*} H_u^\dagger H_u 
    + \eta_{H_d} \frac{X}{M_*} H_d^\dagger H_d 
    + {\rm h.c.} \Biggr) 
\nonumber\\
  && \qquad {} + (\kappa_\Phi)_{ij} \frac{X^\dagger X}{M_*^2} 
      \Phi_i^\dagger \Phi_j 
    + \Biggl( (\eta_\Phi)_{ij} \frac{X}{M_*} \Phi_i^\dagger \Phi_j 
    + {\rm h.c.} \Biggr) \Biggr] 
\nonumber\\
  && {} + \Biggl[ \int\! d^2\theta \Biggl( 
    (\zeta_u)_{ij} \frac{X}{M_*} Q_i U_j H_u 
    + (\zeta_d)_{ij} \frac{X}{M_*} Q_i D_j H_d 
    + (\zeta_e)_{ij} \frac{X}{M_*} L_i E_j H_d \Biggr) 
    + {\rm h.c.} \Biggr],
\label{eq:L_soft-M*}
\end{eqnarray}
where $X = \theta^2 F_X$ is a chiral superfield whose $F$-term 
vacuum expectation value is responsible for supersymmetry breaking, 
${\cal W}^A_\alpha$ ($A=1,2,3$) are the field-strength superfields 
for $U(1)_Y$, $SU(2)_L$ and $SU(3)_C$, and $\Phi = Q,U,D,L$ and $E$. 
The $\kappa_\Phi$ are $3 \times 3$ Hermitian matrices, while $\eta_\Phi$, 
$\zeta_u$, $\zeta_d$ and $\zeta_e$ are general complex $3 \times 3$ 
matrices.  (Here, we have omitted the operators involving $N_i$ because 
in most cases they do not affect our analysis.)

Assuming that suppression factors $\epsilon_{\Phi_i}$, $\epsilon_{H_u}$ 
and $\epsilon_{H_d}$ appear associated with the fields $\Phi_i$, $H_u$ 
and $H_d$, we obtain for the parameters in Eq.~(\ref{eq:L_soft-M*})
\begin{equation}
  \kappa_{H_u} \approx \tilde{\kappa}_{H_u} \epsilon_{H_u}^2, 
\qquad
  \kappa_{H_d} \approx \tilde{\kappa}_{H_d} \epsilon_{H_d}^2,
\label{eq:scal-coeff-1}
\end{equation}
\begin{equation}
  \kappa_\mu \approx \tilde{\kappa}_\mu\, \epsilon_{H_u} \epsilon_{H_d}, 
\qquad
  \kappa_b \approx \tilde{\kappa}_b\, \epsilon_{H_u} \epsilon_{H_d}, 
\qquad
  \eta_{H_u} \approx \tilde{\eta}_{H_u} \epsilon_{H_u}^2, 
\qquad
  \eta_{H_d} \approx \tilde{\eta}_{H_d} \epsilon_{H_d}^2,
\label{eq:scal-coeff-2}
\end{equation}
\begin{equation}
  (\kappa_\Phi)_{ij} \approx \tilde{\kappa}_\Phi\, 
    \epsilon_{\Phi_i} \epsilon_{\Phi_j}, 
\qquad
  (\eta_\Phi)_{ij} \approx \tilde{\eta}_\Phi\, 
    \epsilon_{\Phi_i} \epsilon_{\Phi_j},
\label{eq:scal-coeff-3}
\end{equation}
\begin{equation}
  (\zeta_u)_{ij} \approx \tilde{\zeta}_u\, 
    \epsilon_{Q_i} \epsilon_{U_j} \epsilon_{H_u}, 
\qquad
  (\zeta_d)_{ij} \approx \tilde{\zeta}_d\, 
    \epsilon_{Q_i} \epsilon_{D_j} \epsilon_{H_d}, 
\qquad
  (\zeta_e)_{ij} \approx \tilde{\zeta}_e\, 
    \epsilon_{L_i} \epsilon_{E_j} \epsilon_{H_d},
\label{eq:scal-coeff-4}
\end{equation}
and for the Yukawa couplings
\begin{equation}
  (y_u)_{ij} \approx \tilde{y}_u\, 
    \epsilon_{Q_i} \epsilon_{U_j} \epsilon_{H_u}, 
\qquad
  (y_d)_{ij} \approx \tilde{y}_d\, 
    \epsilon_{Q_i} \epsilon_{D_j} \epsilon_{H_d}, 
\qquad
  (y_e)_{ij} \approx \tilde{y}_e\, 
    \epsilon_{L_i} \epsilon_{E_j} \epsilon_{H_d},
\label{eq:scal-Yukawa-prim}
\end{equation}
where tilde parameters represent the ``natural'' size for the 
couplings without the suppression factors.  For example, if the 
theory is strongly coupled at $M_*$, $\tilde{y}_u \sim \tilde{y}_d 
\sim \tilde{y}_e \sim O(4\pi)$, while if it is weakly coupled, we 
expect $\tilde{y}_u \sim \tilde{y}_d \sim \tilde{y}_e \sim O(1)$. 
Note that $O(1)$ coefficients are omitted in the expressions of 
Eqs.~(\ref{eq:scal-coeff-1}~--~\ref{eq:scal-Yukawa-prim}); for 
example, $(\kappa_\Phi)_{ij}$ is not proportional to $(\eta_\Phi)_{ij}$ 
because of an arbitrary $O(1)$ coefficient in each element.

Depending on the setup, some of the coefficients may be vanishing. 
For example, if the supersymmetry breaking sector does not contain an 
``elementary'' gauge singlet at $M_*$, then $\eta_A = \tilde{\kappa}_\mu 
= \tilde{\eta}_{H_u} = \tilde{\eta}_{H_d} = \tilde{\eta}_\Phi = 
\tilde{\zeta}_u = \tilde{\zeta}_d = \tilde{\zeta}_e = 0$, and the 
gaugino masses must be generated by some low energy mechanism.  (The 
supersymmetric Higgs mass, the $\mu$ term, must also be generated 
at low energies unless it exists at $M_*$ in the superpotential.) 
The precise pattern for $\eta_A$ and the tilde parameters affects 
low energy phenomenology, but our analysis of flavor and $CP$ 
violation is independent of the detailed pattern.

In this paper we consider the case where $\epsilon_{H_u} \sim 
\epsilon_{H_d} \sim O(1)$, and assume for simplicity that the two 
Higgs doublets obey the same scaling, $\tilde{\kappa}_{H_u} \sim 
\tilde{\kappa}_{H_d} \sim \tilde{\kappa}_H$ and $\tilde{\eta}_{H_u} 
\sim \tilde{\eta}_{H_d} \sim \tilde{\eta}_H$, as do the matter fields, 
$\tilde{\kappa}_Q \sim \tilde{\kappa}_U \sim \tilde{\kappa}_D \sim 
\tilde{\kappa}_L \sim \tilde{\kappa}_E \sim \tilde{\kappa}_\Phi$ 
and $\tilde{\eta}_Q \sim \tilde{\eta}_U \sim \tilde{\eta}_D \sim 
\tilde{\eta}_L \sim \tilde{\eta}_E \sim \tilde{\eta}_\Phi$, leading 
to $\tilde{\zeta}_u \sim \tilde{\zeta}_d \sim \tilde{\zeta}_e \sim 
\tilde{\zeta}$ and $\tilde{y}_u \sim \tilde{y}_d \sim \tilde{y}_e 
\sim \tilde{y}$.  An extension to more general cases is straightforward. 
The supersymmetry breaking (and $\mu$) parameters are then obtained 
from Eqs.~(\ref{eq:L_soft-M*}~--~\ref{eq:scal-coeff-4}) as
\begin{equation}
  M_A \approx \eta_A M_{\rm SUSY}, 
\qquad
  \mu \approx \tilde{\kappa}_\mu M_{\rm SUSY}^\dagger, 
\qquad
  b \approx \bigl( \tilde{\kappa}_b 
    + \tilde{\kappa}_\mu \tilde{\eta}_H \bigr) |M_{\rm SUSY}|^2,
\label{eq:gaugino-mu-b}
\end{equation}
\begin{equation}
  m_{H_u}^2 \approx m_{H_d}^2 \approx \bigl( \tilde{\kappa}_H 
    + |\tilde{\eta}_H|^2 \bigr) |M_{\rm SUSY}|^2, 
\qquad
  (m_\Phi^2)_{ij} \approx \{ (\kappa_\Phi)_{ij} 
    + (\eta_\Phi^\dagger \eta_\Phi)_{ij} \} |M_{\rm SUSY}|^2,
\label{eq:soft-masses}
\end{equation}
\begin{equation}
  (a_u)_{ij} 
    \approx \bigl\{ (y_u)_{kj} (\eta_Q)_{ki} + (y_u)_{ik} (\eta_U)_{kj} 
      + (y_u)_{ij} \tilde{\eta}_H \bigr\} M_{\rm SUSY} 
    + \tilde{\zeta}\, \epsilon_{Q_i} \epsilon_{U_j} M_{\rm SUSY},
\label{eq:a_u}
\end{equation}
\begin{equation}
  (a_d)_{ij} 
    \approx \bigl\{ (y_d)_{kj} (\eta_Q)_{ki} + (y_d)_{ik} (\eta_D)_{kj} 
      + (y_d)_{ij} \tilde{\eta}_H \bigr\} M_{\rm SUSY} 
    + \tilde{\zeta}\, \epsilon_{Q_i} \epsilon_{D_j} M_{\rm SUSY},
\label{eq:a_d}
\end{equation}
\begin{equation}
  (a_e)_{ij} 
    \approx \bigl\{ (y_e)_{kj} (\eta_L)_{ki} + (y_e)_{ik} (\eta_E)_{kj} 
      + (y_e)_{ij} \tilde{\eta}_H \bigr\} M_{\rm SUSY} 
    + \tilde{\zeta}\, \epsilon_{L_i} \epsilon_{E_j} M_{\rm SUSY},
\label{eq:a_e}
\end{equation}
where $M_{\rm SUSY} \equiv F_X/M_*$, and $M_A$ are the gaugino masses, 
$m_{H_u}^2$, $m_{H_d}^2$ and $m_\Phi^2$ are non-holomorphic supersymmetry 
breaking squared masses, $b$ is the holomorphic supersymmetry breaking 
Higgs mass-squared, and $(a_u)_{ij}$, $(a_d)_{ij}$ and $(a_e)_{ij}$ 
are holomorphic supersymmetry breaking scalar trilinear interactions. 
We find that the pattern of the supersymmetry breaking parameters 
is correlated with that of the Yukawa couplings, which now read
\begin{equation}
  (y_u)_{ij} \approx \tilde{y}\, \epsilon_{Q_i} \epsilon_{U_j}, 
\qquad
  (y_d)_{ij} \approx \tilde{y}\, \epsilon_{Q_i} \epsilon_{D_j}, 
\qquad
  (y_e)_{ij} \approx \tilde{y}\, \epsilon_{L_i} \epsilon_{E_j}.
\label{eq:scal-Yukawa}
\end{equation}
In general, the correlation between Eqs.~(\ref{eq:gaugino-mu-b}%
~--~\ref{eq:a_e}) and Eq.~(\ref{eq:scal-Yukawa}) significantly 
reduces the tension between supersymmetry breaking and flavor 
physics~\cite{Kitano:2006ws}.  We note again that $O(1)$ 
coefficients are omitted in each term in Eqs.~(\ref{eq:gaugino-mu-b}%
~--~\ref{eq:scal-Yukawa}); for instance, the last terms of 
Eqs.~(\ref{eq:a_u}~--~\ref{eq:a_e}) are not proportional to 
the corresponding Yukawa matrices, Eq.~(\ref{eq:scal-Yukawa}), 
because of these $O(1)$ coefficients.

Taking $\epsilon_{\Phi_1} \leq \epsilon_{\Phi_2} \leq \epsilon_{\Phi_3}$ 
without loss of generality, the Yukawa couplings of Eq.~(\ref{eq:scal-Yukawa}) 
lead to the following quark and lepton masses and mixings
\begin{equation}
\begin{array}{lll}
  (m_t,m_c,m_u) & \approx & \tilde{y}\, \langle H_u \rangle\, 
    (\epsilon_{Q_3}\epsilon_{U_3},\, \epsilon_{Q_2}\epsilon_{U_2},\, 
     \epsilon_{Q_1}\epsilon_{U_1}), \\
  (m_b,m_s,m_d) & \approx & \tilde{y}\, \langle H_d \rangle\, 
    (\epsilon_{Q_3}\epsilon_{D_3},\, \epsilon_{Q_2}\epsilon_{D_2},\, 
     \epsilon_{Q_1}\epsilon_{D_1}), \\
  (m_\tau,m_\mu,m_e) & \approx & \tilde{y}\, \langle H_d \rangle\, 
    (\epsilon_{L_3}\epsilon_{E_3},\, \epsilon_{L_2}\epsilon_{E_2},\, 
     \epsilon_{L_1}\epsilon_{E_1}), \\
  (m_{\nu_\tau},m_{\nu_\mu},m_{\nu_e}) & \approx & 
    \frac{\tilde{y}^2 \langle H_u \rangle^2}{M_N}\, 
    (\epsilon_{L_3}^2,\, \epsilon_{L_2}^2,\, \epsilon_{L_1}^2),
\end{array}
\label{eq:q-l-masses}
\end{equation}
and
\begin{equation}
  V_{\rm CKM} \approx
  \left( \begin{array}{ccc}
    1 & \epsilon_{Q_1}/\epsilon_{Q_2} & \epsilon_{Q_1}/\epsilon_{Q_3} \\
    \epsilon_{Q_1}/\epsilon_{Q_2} & 1 & \epsilon_{Q_2}/\epsilon_{Q_3} \\
    \epsilon_{Q_1}/\epsilon_{Q_3} & \epsilon_{Q_2}/\epsilon_{Q_3} & 1
  \end{array} \right),
\qquad
  V_{\rm MNS} \approx
  \left( \begin{array}{ccc}
    1 & \epsilon_{L_1}/\epsilon_{L_2} & \epsilon_{L_1}/\epsilon_{L_3} \\
    \epsilon_{L_1}/\epsilon_{L_2} & 1 & \epsilon_{L_2}/\epsilon_{L_3} \\
    \epsilon_{L_1}/\epsilon_{L_3} & \epsilon_{L_2}/\epsilon_{L_3} & 1
  \end{array} \right),
\label{eq:q-l-mixings}
\end{equation}
where we have included the neutrino masses through the seesaw 
mechanism with the right-handed neutrino Majorana masses 
$W \approx M_N \epsilon_{N_i} \epsilon_{N_j} N_i N_j$, and 
$V_{\rm CKM}$ and $V_{\rm MNS}$ are the quark and lepton mixing 
matrices, respectively.  The values of the $\epsilon$ parameters 
are then constrained by the observed quark and lepton masses 
and mixings.

There are a variety of possibilities for the origin of the $\epsilon$ 
factors. They may arise, for example, from distributions of fields 
in higher dimensional spacetime or from strong conformal dynamics 
at or above the scale $M_*$.  In a forthcoming paper we will discuss 
an explicit example of such models.  In general, if the suppressions 
of the Yukawa couplings arise from wavefunction effects in a broad 
sense, as in the examples described above, we can obtain the 
correlation given in Eqs.~(\ref{eq:gaugino-mu-b}~--~\ref{eq:a_e}) 
and Eq.~(\ref{eq:scal-Yukawa}).  Another possibility is to introduce 
a non-Abelian flavor symmetry connecting all three generations. 
Flavor violating supersymmetry breaking parameters having a similar 
correlation to Eqs.~(\ref{eq:gaugino-mu-b}~--~\ref{eq:scal-Yukawa}) 
may then be generated through the breaking of that symmetry.%
\footnote{For earlier analyses on flavor violation in models with 
 non-Abelian flavor symmetries, see e.g.~\cite{Pomarol:1995xc}.}
While this allows flavor universal contributions to the supersymmetry 
breaking parameters in addition to Eqs.~(\ref{eq:gaugino-mu-b}~--%
~\ref{eq:a_e}), the essential features of the framework are not 
affected.

\section{Constraints from Low Energy Processes}
\label{sec:constraint}

The supersymmetry breaking parameters are subject to a number 
of constraints from low energy flavor and $CP$ violating processes. 
Here we study these constraints for the parameters given in 
Eqs.~(\ref{eq:gaugino-mu-b}~--~\ref{eq:a_e}).  We assume that 
$CP$ violating effects associated with the Higgs sector are 
sufficiently suppressed.  This is achieved if either $b \ll |\mu|^2$ 
at $M_*$ or the phases of $\mu$ and $b$ are aligned in the basis 
where the $M_A$ are real.

The values of low energy supersymmetry breaking parameters are obtained 
from Eqs.~(\ref{eq:gaugino-mu-b}~--~\ref{eq:a_e}) by evolving them down 
to the weak scale using renormalization group equations.  Contributions 
from other flavor universal sources, such as gauge mediation, may 
also be added.  To parameterize these effects in a model-independent 
manner, we simply add universal squark and slepton squared masses, 
$m_{\tilde{q}}^2 \equiv \lambda_{\tilde{q}}^2\, |M_{\rm SUSY}|^2$ 
and $m_{\tilde{l}}^2 \equiv \lambda_{\tilde{l}}^2\, |M_{\rm SUSY}|^2$, 
to $(m_\Phi^2)_{ij}$:
\begin{equation}
  (m_\Phi^2)_{ij} 
  \rightarrow \left\{
  \begin{array}{ll}
    (m_\Phi^2)_{ij} + \lambda_{\tilde{q}}^2\, 
      |M_{\rm SUSY}|^2 \delta_{ij} \, & {\rm for}\,\,\, \Phi = Q, U, D 
\\
    (m_\Phi^2)_{ij} + \lambda_{\tilde{l}}^2\, 
      |M_{\rm SUSY}|^2 \delta_{ij} \, & {\rm for}\,\,\, \Phi = L, E 
  \end{array}. \right.
\label{eq:m2-LE}
\end{equation}
We neglect the differences of the flavor universal contribution among 
various squarks and among various sleptons, but it is sufficient 
for our purposes here.  Effects on the gaugino masses and the scalar 
trilinear interactions are absorbed into the redefinition of $\eta_A$ 
and $\tilde{\eta}_H$, respectively.  The $m_{H_u}^2$ and $m_{H_d}^2$ 
are also renormalized, but this effect is incorporated by treating 
$\tan\beta \equiv \langle H_u \rangle/\langle H_d \rangle$ as 
a free parameter.

We use the mass insertion method~\cite{Hall:1985dx} to compare the 
expected amount of flavor violation in the present scenario to low 
energy data. In order to do a mass insertion analysis, we need to 
work in the super-CKM basis where the Yukawa matrices are diagonalized 
by supersymmetric rotations of $Q_i$, $U_i$, $D_i$, $L_i$ and $E_i$. 
The mass insertion parameters, $\delta_{ij}$, are then obtained by 
dividing the off-diagonal entry of the sfermion mass-squared matrix 
by the average diagonal entry.  Using Eqs.~(\ref{eq:gaugino-mu-b}%
~--~\ref{eq:a_e},~\ref{eq:m2-LE}), we obtain
\begin{equation}
  (\delta^u_{ij})_{LL} \approx \frac{1}{\lambda_{\tilde{q}}^2} 
    \Bigl( \tilde{\kappa}_\Phi + |\tilde{\eta}_\Phi|^2 
      \epsilon_{Q_3}^2 \Bigr)\, \epsilon_{Q_i} \epsilon_{Q_j},
\qquad
  (\delta^u_{ij})_{RR} \approx \frac{1}{\lambda_{\tilde{q}}^2} 
    \Bigl( \tilde{\kappa}_\Phi + |\tilde{\eta}_\Phi|^2 
      \epsilon_{U_3}^2 \Bigr)\, \epsilon_{U_i} \epsilon_{U_j},
\label{eq:delta_u_LL-RR}
\end{equation}
\begin{equation}
  (\delta^u_{ij})_{LR} = (\delta^u_{ji})_{RL}^* 
  \approx \frac{1}{\lambda_{\tilde{q}}^2} 
    \Bigl\{ \tilde{y}\, \tilde{\eta}_\Phi 
        (\epsilon_{Q_j}^2 + \epsilon_{U_i}^2) 
      + \tilde{\zeta} \Bigr\}\, \epsilon_{Q_i} \epsilon_{U_j} 
    \frac{v \sin\beta}{M_{\rm SUSY}},
\label{eq:delta_u_LR-RL}
\end{equation}
for the up-type squarks,
\begin{equation}
  (\delta^d_{ij})_{LL} \approx \frac{1}{\lambda_{\tilde{q}}^2} 
    \Bigl( \tilde{\kappa}_\Phi + |\tilde{\eta}_\Phi|^2 
      \epsilon_{Q_3}^2 \Bigr)\, \epsilon_{Q_i} \epsilon_{Q_j},
\qquad
  (\delta^d_{ij})_{RR} \approx \frac{1}{\lambda_{\tilde{q}}^2} 
    \Bigl( \tilde{\kappa}_\Phi + |\tilde{\eta}_\Phi|^2 
      \epsilon_{D_3}^2 \Bigr)\, \epsilon_{D_i} \epsilon_{D_j},
\label{eq:delta_d_LL-RR}
\end{equation}
\begin{equation}
  (\delta^d_{ij})_{LR} = (\delta^d_{ji})_{RL}^* 
  \approx \frac{1}{\lambda_{\tilde{q}}^2} 
    \Bigl\{ \tilde{y}\, \tilde{\eta}_\Phi 
        (\epsilon_{Q_j}^2 + \epsilon_{D_i}^2) 
      + \tilde{\zeta} \Bigr\}\, \epsilon_{Q_i} \epsilon_{D_j} 
    \frac{v \cos\beta}{M_{\rm SUSY}},
\label{eq:delta_d_LR-RL}
\end{equation}
for the down-type squarks,
\begin{equation}
  (\delta^e_{ij})_{LL} \approx \frac{1}{\lambda_{\tilde{l}}^2} 
    \Bigl( \tilde{\kappa}_\Phi + |\tilde{\eta}_\Phi|^2 
      \epsilon_{L_3}^2 \Bigr)\, \epsilon_{L_i} \epsilon_{L_j},
\qquad
  (\delta^e_{ij})_{RR} \approx \frac{1}{\lambda_{\tilde{l}}^2} 
    \Bigl( \tilde{\kappa}_\Phi + |\tilde{\eta}_\Phi|^2 
      \epsilon_{E_3}^2 \Bigr)\, \epsilon_{E_i} \epsilon_{E_j},
\label{eq:delta_e_LL-RR}
\end{equation}
\begin{equation}
  (\delta^e_{ij})_{LR} = (\delta^e_{ji})_{RL}^* 
  \approx \frac{1}{\lambda_{\tilde{l}}^2} 
    \Bigl\{ \tilde{y}\, \tilde{\eta}_\Phi 
        (\epsilon_{L_j}^2 + \epsilon_{E_i}^2) 
      + \tilde{\zeta} \Bigr\}\, \epsilon_{L_i} \epsilon_{E_j} 
    \frac{v \cos\beta}{M_{\rm SUSY}},
\label{eq:delta_e_LR-RL}
\end{equation}
for the charged sleptons, and
\begin{equation}
  (\delta^\nu_{ij})_{LL} \approx \frac{1}{\lambda_{\tilde{l}}^2} 
    \Bigl( \tilde{\kappa}_\Phi + |\tilde{\eta}_\Phi|^2 
      \epsilon_{L_3}^2 \Bigr)\, \epsilon_{L_i} \epsilon_{L_j},
\label{eq:delta_nu_LL}
\end{equation}
for the sneutrinos.  Here, $v \equiv (\langle H_u \rangle^2 
+ \langle H_d \rangle^2)^{1/2} \simeq 174~{\rm GeV}$ and 
$\tan\beta = \langle H_u \rangle/\langle H_d \rangle$.

The values of the $\epsilon$ parameters are constrained to 
reproduce the observed quark and lepton masses and mixings through 
Eqs.~(\ref{eq:q-l-masses},~\ref{eq:q-l-mixings}).  They depend on 
$\tilde{y}$ as well as the value of $\tan\beta$.  For illustrative 
purpose, we take the pattern
\begin{equation}
\begin{array}{lll}
  \epsilon_{Q_1} \approx 
    \tilde{y}^{-\frac{1}{2}} \alpha_q\, \epsilon^2,                 \quad & 
  \epsilon_{U_1} \approx 
    \tilde{y}^{-\frac{1}{2}} \alpha_q^{-1} \epsilon^2,              \quad & 
  \epsilon_{D_1} \approx 
    \tilde{y}^{-\frac{1}{2}} \alpha_q^{-1} \alpha_\beta\, \epsilon,
\\
  \epsilon_{Q_2} \approx 
    \tilde{y}^{-\frac{1}{2}} \alpha_q\, \epsilon,                   \quad & 
  \epsilon_{U_2} \approx 
    \tilde{y}^{-\frac{1}{2}} \alpha_q^{-1} \epsilon,                \quad & 
  \epsilon_{D_2} \approx 
    \tilde{y}^{-\frac{1}{2}} \alpha_q^{-1} \alpha_\beta\, \epsilon,
\\
  \epsilon_{Q_3} \approx 
    \tilde{y}^{-\frac{1}{2}} \alpha_q,                              \quad & 
  \epsilon_{U_3} \approx 
    \tilde{y}^{-\frac{1}{2}} \alpha_q^{-1},                         \quad & 
  \epsilon_{D_3} \approx 
    \tilde{y}^{-\frac{1}{2}} \alpha_q^{-1} \alpha_\beta\, \epsilon,
\end{array}
\label{eq:epsilon-QUD}
\end{equation}
\begin{equation}
\begin{array}{ll}
  \epsilon_{L_1} \approx 
    \tilde{y}^{-\frac{1}{2}} \alpha_l\, \epsilon,                     \quad & 
  \epsilon_{E_1} \approx 
    \tilde{y}^{-\frac{1}{2}} \alpha_l^{-1} \alpha_\beta\, \epsilon^2,
\\
  \epsilon_{L_2} \approx 
    \tilde{y}^{-\frac{1}{2}} \alpha_l\, \epsilon,                     \quad & 
  \epsilon_{E_2} \approx 
    \tilde{y}^{-\frac{1}{2}} \alpha_l^{-1} \alpha_\beta\, \epsilon,
\\
  \epsilon_{L_3} \approx 
    \tilde{y}^{-\frac{1}{2}}  \alpha_l\, \epsilon,                    \quad & 
  \epsilon_{E_3} \approx 
    \tilde{y}^{-\frac{1}{2}} \alpha_l^{-1} \alpha_\beta,
\end{array}
\label{eq:epsilon-LE}
\end{equation}
with
\begin{equation}
  \tan\beta \approx \alpha_\beta\, \epsilon^{-1},
\label{eq:tan-beta}
\end{equation}
where $\epsilon \sim O(0.1)$ and $\alpha_q$, $\alpha_l$ and $\alpha_\beta$ 
are numbers parameterizing the freedoms unfixed by the data of the quark 
and lepton masses and mixings.  Here, we have assumed that $\tan\beta$ 
is larger than $\approx 2$, as suggested by the large top quark mass. 
The pattern of Eq.~(\ref{eq:epsilon-QUD}~--~\ref{eq:tan-beta}) leads to
\begin{equation}
\begin{array}{lll}
  (m_t,m_c,m_u) & \approx & v\, (1,\epsilon^2,\epsilon^4),               \\
  (m_b,m_s,m_d) & \approx & v\, (\epsilon^2,\epsilon^3,\epsilon^4),      \\
  (m_\tau,m_\mu,m_e) & \approx & v\, (\epsilon^2,\epsilon^3,\epsilon^4), \\
  (m_{\nu_\tau},m_{\nu_\mu},m_{\nu_e}) & \approx & \frac{v^2}{M_N} (1,1,1),
\end{array}
\label{eq:q-l-masses-2}
\end{equation}
and
\begin{equation}
  V_{\rm CKM} \approx
  \left( \begin{array}{ccc}
    1 & \epsilon & \epsilon^2 \\
    \epsilon & 1 & \epsilon   \\
    \epsilon^2 & \epsilon & 1
  \end{array} \right),
\qquad
  V_{\rm MNS} \approx
  \left( \begin{array}{ccc}
    1 & 1 & 1 \\
    1 & 1 & 1 \\
    1 & 1 & 1
  \end{array} \right),
\label{eq:q-l-mixings-2}
\end{equation}
which successfully reproduces the gross structure of the observed quark 
and lepton masses and mixings~\cite{Hall:1999sn}.  The mass insertion 
parameters are obtained by substituting Eqs.~(\ref{eq:epsilon-QUD}%
~--~\ref{eq:tan-beta}) into Eqs.~(\ref{eq:delta_u_LL-RR}%
~--~\ref{eq:delta_nu_LL}).

Here we summarize the constraints from low energy flavor and $CP$ 
violating processes, compiled from Ref.~\cite{Gabbiani:1996hi}. In the 
quark sector, the most stringent experimental constraints come from 
$K$-$\bar{K}$, $D$-$\bar{D}$ and $B$-$\bar{B}$ mixings, $\sin 2\beta$ 
and the $b \rightarrow s\gamma$ process.  The model-independent 
constraints are obtained by turning on only one (or two) mass insertion 
parameter(s) and considering the gluino exchange diagrams.  They 
are summarized as
\begin{eqnarray}
\begin{array}{lll}
  \sqrt{|{\rm Re}(\delta^d_{12})_{LL/RR}^2|} 
    \simlt (10^{-2}\mbox{--}10^{-1}), & 
  \sqrt{|{\rm Re}(\delta^d_{12})_{LR/RL}^2|} 
    \simlt (10^{-3}\mbox{--}10^{-2}), & 
  \sqrt{|{\rm Re}(\delta^d_{12})_{LL}(\delta^d_{12})_{RR}|} \simlt 10^{-3},
\nonumber\\ \\
  \sqrt{|{\rm Im}(\delta^d_{12})_{LL/RR}^2|} 
    \simlt (10^{-3}\mbox{--}10^{-2}), & 
  \sqrt{|{\rm Im}(\delta^d_{12})_{LR/RL}^2|} 
    \simlt (10^{-4}\mbox{--}10^{-3}), & 
  \sqrt{|{\rm Im}(\delta^d_{12})_{LL}(\delta^d_{12})_{RR}|} \simlt 10^{-4},
\end{array}
\nonumber
\label{eq:delta-exp_KK}
\end{eqnarray}
\begin{eqnarray}
\begin{array}{lll}
  |(\delta^u_{12})_{LL/RR}| \simlt (10^{-2}\mbox{--}10^{-1}), & 
  |(\delta^u_{12})_{LR/RL}| \simlt 10^{-2}, & 
  |(\delta^u_{12})_{LL}| = |(\delta^u_{12})_{RR}| 
    \simlt (10^{-3}\mbox{--}10^{-2}),
\nonumber\\ \\
  |(\delta^d_{13})_{LL/RR}| \simlt (0.1\mbox{--}1), & 
  |(\delta^d_{13})_{LR/RL}| \simlt (10^{-2}\mbox{--}10^{-1}), & 
  |(\delta^d_{13})_{LL}| = |(\delta^d_{13})_{RR}| \simlt 10^{-2},
\end{array}
\nonumber
\label{eq:delta-exp_DD-BB}
\end{eqnarray}
\begin{eqnarray}
  |(\delta^d_{23})_{LR/RL}| \simlt 10^{-2},
\nonumber
\label{eq:delta-exp_bs}
\end{eqnarray}
where we have taken the gluino and squark masses to be the same 
order of magnitude and $m_{\tilde{q}} = 500~{\rm GeV}$.  For heavier 
superparticles, the bounds scale roughly linearly with $m_{\tilde{q}}$, 
i.e. all the bounds weaken for larger $m_{\tilde{q}}$ by a factor of 
$m_{\tilde{q}}/500~{\rm GeV}$.  In the lepton sector, the most stringent 
constraint comes from the $\mu \rightarrow e \gamma$ process, and is 
given by
\begin{eqnarray}
\begin{array}{ll}
  |(\delta^e_{12})_{LL}| \simlt (10^{-4}\mbox{--}10^{-3}),\quad & 
  |(\delta^e_{12})_{LR/RL}| \simlt (10^{-6}\mbox{--}10^{-5}),
\end{array}
\nonumber
\label{eq:delta-exp_LFV}
\end{eqnarray}
where we have taken the weak gaugino and slepton masses to be the same 
order of magnitude and $m_{\tilde{l}} = 200~{\rm GeV}$.  For heavier 
superparticles, the bound on $|(\delta^e_{12})_{LL}|$ scales roughly 
quadratically with $m_{\tilde{l}}$, while that on $|(\delta^e_{12})_{LR/RL}|$ 
scales roughly linearly with $m_{\tilde{l}}$.  Finally, the bounds from 
the neutron and electron electric dipole moments (EDMs) constrain the 
flavor conserving entry of the sfermion mass matrices:
\begin{eqnarray}
\begin{array}{lll}
  |{\rm Im}(\delta^u_{11})_{LR}| \simlt 10^{-6},\quad & 
  |{\rm Im}(\delta^d_{11})_{LR}| \simlt 10^{-6},\quad & 
  |{\rm Im}(\delta^e_{11})_{LR}| \simlt 10^{-7},
\end{array}
\nonumber
\label{eq:delta-exp_EDM}
\end{eqnarray}
where we have again taken $m_{\tilde{q}} = 500~{\rm GeV}$ and 
$m_{\tilde{l}} = 200~{\rm GeV}$, and the bounds become weaker linearly 
with increasing superparticle masses.

We now determine whether flavor and $CP$ violation arising from 
Eqs.~(\ref{eq:delta_u_LL-RR}~--~\ref{eq:tan-beta}) is compatible 
with the experimental bounds given above.  We take $\epsilon \approx 
(0.05$~--~$0.1)$ to reproduce the gross structure of the quark 
and lepton masses and mixings, and take $\tilde{\kappa}_\Phi \sim 
\tilde{\eta}_\Phi \sim O(1)$ for simplicity.  For $\tilde{\zeta} 
\sim O(1)$, we find stringent constraints coming from the electron 
EDM and $\mu \rightarrow e\gamma$, which push the supersymmetry 
breaking scale up to $M_{\rm SUSY} \simgt 5~{\rm TeV}$ for $\tilde{y} 
\sim 1$ and $M_{\rm SUSY} \simgt 1.5~{\rm TeV}$ for $\tilde{y} \sim 
4\pi$.  Here, we have taken $M_{\rm SUSY} \approx m_{\tilde{q}} \approx 
(5/2) m_{\tilde{l}}$.  This implies that the superpotential couplings 
in Eq.~(\ref{eq:L_soft-M*}), $\tilde{\zeta}$, must somehow be suppressed, 
unless the superparticles are relatively heavy.  This may naturally 
arise from physics above $M_*$, since the superpotential has the special 
property of not being renormalized at all orders in perturbation theory. 

For $\tilde{\zeta} \ll 1$, a wide parameter region is open.  For 
$\tilde{y} \sim 1$, we find that the region
\begin{equation}
  0.2 \simlt \alpha_q \simlt 3,
\qquad
  \frac{\alpha_q}{\alpha_\beta} \simgt 0.5,
\qquad
  \alpha_l \simlt 0.3,
\qquad
  \frac{\alpha_l}{\alpha_\beta} \simgt 0.2,
\label{eq:region-1}
\end{equation}
satisfies all the constraints for $\epsilon = 0.05$, $M_{\rm SUSY} 
= m_{\tilde{q}} = 500~{\rm GeV}$ and $m_{\tilde{l}} = 200~{\rm GeV}$. 
The region is somewhat smaller for $\epsilon = 0.1$.  For $\tilde{y} 
\sim 4\pi$, we find the region
\begin{equation}
  0.05 \simlt \alpha_q \simlt 10,
\qquad
  \frac{\alpha_q}{\alpha_\beta} \simgt 0.1,
\qquad
  \alpha_l \simlt 1,
\qquad
  \frac{\alpha_l}{\alpha_\beta} \simgt 0.04,
\label{eq:region-2}
\end{equation}
for $\epsilon = 0.05$, $M_{\rm SUSY} = m_{\tilde{q}} = 500~{\rm GeV}$ and 
$m_{\tilde{l}} = 200~{\rm GeV}$, and somewhat smaller for $\epsilon = 0.1$.%
\footnote{If we require the absence of cancellation among diagrams for 
 $\mu \rightarrow e \gamma$, the bound $|(\delta^e_{12})_{RR}| \simlt 
 (10^{-3}\mbox{--}10^{-2})$ arises.  This, however, changes the regions 
 of Eqs.~(\ref{eq:region-1},~\ref{eq:region-2}) only slightly.  The lower 
 bound on $\alpha_l/\alpha_\beta$ becomes $0.5$ in Eq.~(\ref{eq:region-1}), 
 and $0.1$ in Eq.~(\ref{eq:region-2}).}
This result agrees with that of Ref.~\cite{Kitano:2006ws}, 
which analyzed the case of $\tilde{y} \sim 1$ without including the 
constraints from the EDM bounds.

We have used the particular parameterization of Eq.~(\ref{eq:epsilon-QUD}%
~--~\ref{eq:tan-beta}) in the analysis here, but we can adopt a 
more refined scheme for the values of the $\epsilon$ parameters to 
better accommodate the observed quark and lepton masses and mixings. 
For example, we can make $\epsilon_{L_1}$ somewhat smaller than 
Eq.~(\ref{eq:epsilon-LE}) to explain the smallness of the $e3$ element 
of $V_{\rm MNS}$, which is experimentally smaller than about $0.2$. 
A value of $\tan\beta$ somewhat larger than Eq.~(\ref{eq:tan-beta}) 
also improves the top to bottom mass ratio.  Our basic results above 
are not affected by these modifications.

We conclude that the current experimental constraints allow the 
existence of the general supersymmetry breaking parameters of 
Eq.~(\ref{eq:L_soft-M*}) where the couplings are suppressed by 
the factors suggested by Yukawa couplings, as long as the superparticles 
are relatively heavy or the superpotential couplings, $\tilde{\zeta}$, 
are suppressed.  For $\tilde{\zeta} \ll 1$, a wide parameter region 
is open even for light superparticles, $m_{\tilde{q}} \approx 
500~{\rm GeV}$ and $m_{\tilde{l}} \approx 200~{\rm GeV}$.  In the 
next section, we discuss implications of this scenario, which we 
call flavorful supersymmetry, on the low energy spectrum.

\section{Implications on the Superparticle Spectrum}
\label{sec:implication}

Phenomenology of supersymmetric theories depends strongly on the 
spectrum of superparticles.  In particular, the order of the superparticle 
masses controls decay chains, and thus affects collider signatures 
significantly.  In this section we discuss the splitting and ordering 
of the superparticle masses among different generations and between 
different superparticle species. 

\subsection{Mass splitting and ordering among generations}
\label{subsec:ordering}

Among the various sfermions, the lightest species are most likely the 
right-handed sleptons: $\tilde{e}_R$, $\tilde{\mu}_R$ and $\tilde{\tau}_R$. 
This is because the sfermion squared masses receive positive contributions 
from the gaugino masses through renormalization group evolution at 
one loop, and these contributions are proportional to the square of 
the relevant gauge couplings.  Since the right-handed sleptons are 
charged under only $U(1)_Y$, they receive contributions from just 
the hypercharge gaugino and are expected to be lighter than the 
other sfermions.  A possible low energy gauge mediated contribution 
will not change the situation because it gives positive contributions 
to the sfermion squared masses proportional to the fourth power 
of the relevant gauge couplings (at least in the simplest case). 
Thus we focus on the right-handed sleptons and analyze the mass 
splitting among the three generations.  A similar analysis, however, 
can also be performed for the other sfermion species.%
\footnote{The mass splitting and ordering for heavier species 
 may also provide important tests for flavorful supersymmetry. 
 Moreover, if there exists a $U(1)_Y$ $D$-term contribution, i.e. 
 $m_{H_u}^2 - m_{H_d}^2 + {\rm Tr}[m_Q^2 -2 m_U^2 + m_D^2 - m_L^2 
 + m_E^2] \neq 0$, then the left-handed sleptons and sneutrinos 
 may be lighter than the right-handed sleptons.  It is also possible 
 to consider the case in which a squark is the lightest sfermion 
 if $M_3$ is significantly smaller than $M_{1,2}$ at $M_*$.}

We consider the field basis in which the lepton Yukawa couplings, 
$(y_e)_{ij}$, are real and diagonal.  If there is no intrinsic flavor 
violation in the sfermion masses, the $3 \times 3$ mass-squared matrix 
for the right-handed sleptons, $m_E^2$, receives a flavor universal 
contribution, $m_{\tilde{e}}^2\,{\rm diag}(1,1,1)$, and flavor 
dependent contributions through renormalization group evolution. 
This leads to
\begin{equation}
  m_E^2 = \left( \begin{array}{ccc}
    m_{\tilde{e}}^2-I_e  &  0 & 0 \\
    0 & m_{\tilde{e}}^2-I_\mu & 0 \\
    0 & 0 & m_{\tilde{e}}^2-I_\tau
  \end{array} \right),
\label{eq:mE2}
\end{equation}
at the weak scale, where $I_e$, $I_\mu$ and $I_\tau$ parameterize 
the effect of renormalization group evolution from the Yukawa and 
scalar trilinear couplings, and $I_e : I_\mu : I_\tau \approx 
(y_e)_{11}^2 : (y_e)_{22}^2 : (y_e)_{33}^2 \approx m_e^2 : m_\mu^2 
: m_\tau^2$.  (The effects from the neutrino Yukawa couplings that 
may exist above the scale of right-handed neutrino masses, $M_N$, are 
neglected here since they are expected to be small.)  The expression 
of Eq.~(\ref{eq:mE2}) tells us that, in the absence of a flavor 
violating contribution, (i) the interaction and mass eigenstates 
of the right-handed sleptons coincide, and (ii) the mass of a slepton 
corresponding to a heavier lepton is always lighter, since $I_\tau 
> I_\mu > I_e > 0$ due to the form of the renormalization group 
equations of the supersymmetric standard model when $(m_E^2)_{ii}, 
(m_L^2)_{ii}, m_{H_d}^2 > 0$.  Inclusion of flavor universal 
left-right mixing does not change these conclusions.

The situation is very different if there exists intrinsic flavor 
violation in supersymmetry breaking.  The supersymmetry breaking 
parameters at $M_*$ in our scenario are given parametrically by 
Eqs.~(\ref{eq:gaugino-mu-b}~--~\ref{eq:a_e},~\ref{eq:scal-coeff-3}) 
even in the basis where $(y_e)_{ij}$ is diagonal.  In particular,
\begin{equation}
  m_E^2(M_*) \approx \left( \begin{array}{ccc}
    \epsilon_{E_1}^2 & \epsilon_{E_1}\epsilon_{E_2}
      & \epsilon_{E_1}\epsilon_{E_3} \\
    \epsilon_{E_1}\epsilon_{E_2} & \epsilon_{E_2}^2 
      & \epsilon_{E_2}\epsilon_{E_3} \\
    \epsilon_{E_1}\epsilon_{E_3} & \epsilon_{E_2}\epsilon_{E_3} 
      & \epsilon_{E_3}^2
  \end{array} \right) |M_{\rm SUSY}|^2,
\label{eq:mE2-M*}
\end{equation}
for $\tilde{\kappa}_\Phi \sim \tilde{\eta}_\Phi \sim O(1)$.  In addition, 
$m_E^2$ receives universal contributions from the $U(1)_Y$ gaugino mass 
through renormalization group evolution, as well as possibly from other 
sources such as low energy gauge mediation.  It also receives flavor 
violating contributions from the Yukawa and scalar trilinear couplings 
through renormalization group evolution.  We find that the evolution 
effect on the off-diagonal elements is not significant; the changes 
of the coefficients are at most of order unity.  The diagonal elements 
receive flavor universal contributions, which we denote as $m_{\tilde{e}}^2 
\equiv \xi_{\tilde{e}}^2\, |M_{\rm SUSY}|^2$, as well as flavor 
dependent contributions.  Defining the flavor dependent part as 
$\hat{m}_{E_i}^2 \equiv (m_E^2)_{ii} - (m_E^2)_{ii}|_{y_e = a_e = 0}$, 
the renormalization group equation for $\hat{m}_{E_i}^2$ is given by
\begin{equation}
  \frac{d}{d\ln\mu_R} \hat{m}_{E_i}^2 = \frac{1}{4\pi^2} 
    \left[ (y_e)_{ii}^2 \left\{ (m_E^2)_{ii} + (m_L^2)_{ii} 
      + m_{H_d}^2 \right\} + \sum_k |(a_e)_{ki}|^2 \right],
\label{eq:RGE-mE2}
\end{equation}
where $i$ in the right-hand-side is not summed.  This leads to $m_E^2$ 
at the weak scale of the form
\begin{equation}
  m_E^2 \approx \left( \begin{array}{ccc}
    m_{\tilde{e}}^2-K_e  &  0 & 0 \\
    0 & m_{\tilde{e}}^2-K_\mu & 0 \\
    0 & 0 & m_{\tilde{e}}^2-K_\tau
  \end{array} \right)
  + \left( \begin{array}{ccc}
    \epsilon_{E_1}^2 & \epsilon_{E_1}\epsilon_{E_2}
      & \epsilon_{E_1}\epsilon_{E_3} \\
    \epsilon_{E_1}\epsilon_{E_2} & \epsilon_{E_2}^2 
      & \epsilon_{E_2}\epsilon_{E_3} \\
    \epsilon_{E_1}\epsilon_{E_3} & \epsilon_{E_2}\epsilon_{E_3} 
      & \epsilon_{E_3}^2
  \end{array} \right) |M_{\rm SUSY}|^2,
\label{eq:mE2-fin-1}
\end{equation}
where $O(1)$ coefficients are omitted in each element in the second 
term, but not in the first term.  The quantities $K_e$, $K_\mu$ 
and $K_\tau$ are defined by $K_\tau \equiv \hat{m}_{E_3}^2(M_*) - 
\hat{m}_{E_3}^2(M_{\rm SUSY})$ and $\{\tau,3\} \rightarrow \{e,1\}, 
\{\mu,2\}$, and are given by solving Eq.~(\ref{eq:RGE-mE2}).  They 
are always positive for $(m_E^2)_{ii}, (m_L^2)_{ii}, m_{H_d}^2 > 0$, 
and $K_e : K_\mu : K_\tau \approx (y_e)_{11}^2 : (y_e)_{22}^2 : 
(y_e)_{33}^2$ for $(a_e)_{ij} \propto (y_e)_{ij}$.

The contributions $K_e$, $K_\mu$ and $K_\tau$ compete in general with 
the second term in Eq.~(\ref{eq:mE2-fin-1}).  For $\tilde{\kappa}_\Phi 
\sim \tilde{\eta}_\Phi \sim \tilde{\eta}_H \sim O(1)$ and $\tilde{\zeta} 
\simlt \tilde{y}$, for example, Eq.~(\ref{eq:RGE-mE2}) scales as
\begin{equation}
  \frac{d}{d\ln\mu_R} \hat{m}_{E_i}^2 
    \approx \frac{1}{4\pi^2} (y_e)_{ii}^2 
      \left( \xi_{\tilde{e}}^2 + 2 \xi_{\tilde{l}}^2 
        + \sum_k \frac{\epsilon_{L_k}^2}{\epsilon_{L_i}^2} 
      \right) |M_{\rm SUSY}|^2,
\label{eq:RGE-mE2-approx}
\end{equation}
where we have set $(m_L^2)_{ii} \approx m_{H_d}^2 \equiv 
\xi_{\tilde{l}}^2\, |M_{\rm SUSY}|^2$.  With the choice of 
Eqs.~(\ref{eq:epsilon-QUD}~--~\ref{eq:tan-beta}), this gives
\begin{equation}
  K_\tau \approx \frac{1}{4\pi^2} (y_e)_{33}^2 
      \left( \xi_{\tilde{e}}^2 + 2 \xi_{\tilde{l}}^2 + O(1) \right) 
      |M_{\rm SUSY}|^2 \ln\frac{M_*}{M_{\rm SUSY}}
    \sim \tilde{y}^2 \epsilon_{L_3}^2 \epsilon_{E_3}^2 |M_{\rm SUSY}|^2,
\label{eq:K_e}
\end{equation}
and $\{\tau,3\} \rightarrow \{e,1\}, \{\mu,2\}$, leading to
\begin{equation}
  m_E^2 \approx \left( \begin{array}{ccc}
    \xi_{\tilde{e}}^2 - \tilde{y}^2\epsilon_{L_1}^2\epsilon_{E_1}^2 
      + \epsilon_{E_1}^2 & \epsilon_{E_1}\epsilon_{E_2}
      & \epsilon_{E_1}\epsilon_{E_3} \\
    \epsilon_{E_1}\epsilon_{E_2} & \xi_{\tilde{e}}^2 
      - \tilde{y}^2\epsilon_{L_2}^2\epsilon_{E_2}^2 + \epsilon_{E_2}^2 
      & \epsilon_{E_2}\epsilon_{E_3} \\
    \epsilon_{E_1}\epsilon_{E_3} & \epsilon_{E_2}\epsilon_{E_3} 
      & \xi_{\tilde{e}}^2 - \tilde{y}^2\epsilon_{L_3}^2\epsilon_{E_3}^2 
      + \epsilon_{E_3}^2 
  \end{array} \right) |M_{\rm SUSY}|^2.
\label{eq:mE2-fin-2}
\end{equation}
Note that the signs of the $\tilde{y}^2\epsilon_{L_i}^2\epsilon_{E_i}^2$ 
terms are all negative, while each $\epsilon_{E_i}\epsilon_{E_j}$ term 
has an $O(1)$ coefficient whose sign can be either positive or negative.

The expression of Eq.~(\ref{eq:mE2-fin-2}) shows that in flavorful 
supersymmetry (i) the interaction and mass eigenstates of the 
right-handed sleptons do not in general coincide, and (ii) the mass 
ordering of the sleptons is not necessarily anticorrelated with that 
of the leptons.  In particular, we find that the lightest sfermion 
can easily be $\tilde{e}_R$ or $\tilde{\mu}_R$ (with slight mixtures 
from other flavors), in contrast to the usual supersymmetry breaking 
scenarios in which $\tilde{\tau}_R$ is the lightest because $I_\tau > 
I_\mu > I_e > 0$ in Eq.~(\ref{eq:mE2}).  In our case, $\tilde{\tau}_R$ 
is heavier than $\tilde{e}_R$ and $\tilde{\mu}_R$ if, for example, 
$\tilde{y} \sim 1$ and the $\epsilon_{E_3}^2$ term in the $3$-$3$ 
entry of Eq.~(\ref{eq:mE2-fin-2}) has a positive coefficient.  As 
we will see in section~\ref{sec:LHC}, this can lead to distinct 
signatures at the LHC which provide strong evidence for the present 
scenario.  Note that even when the mass ordering is not flipped, the 
amount of mass splitting between the generations differs from the 
conventional scenarios, which may provide a direct test of this 
scenario at future colliders.  In particular, with our choice of 
Eqs.~(\ref{eq:epsilon-QUD}~--~\ref{eq:tan-beta}), the flavor dependent 
contribution to the $3$-$3$ entry of Eq.~(\ref{eq:mE2-fin-2}), 
$\epsilon_{E_3}^2 |M_{\rm SUSY}|^2$, can be of the same order 
as the flavor universal contributions.  This implies that the 
$\tilde{\tau}_R$ mass may be significantly split from those of 
$\tilde{e}_R$ and $\tilde{\mu}_R$, giving a window into the effect 
of intrinsic flavor violation in the supersymmetry breaking sector. 
The mass splitting between $\tilde{e}_R$ and $\tilde{\mu}_R$ is 
of order $\epsilon_{E_2}^2 |M_{\rm SUSY}|^2$, which can also be 
much larger than the conventional scenarios and may be measurable.

\subsection{The lightest and next-to-lightest supersymmetric particles}
\label{subsec:LSP-NLSP}

Phenomenology at colliders depends strongly on the species of the 
lightest superparticle (LSP) and the next-to-lightest superparticle 
(NLSP).  As we have seen, it is natural to expect that (any) one 
of the right-handed sleptons is the lightest sfermion.  For the 
gauginos, we expect that the bino, $\tilde{B}$, is naturally the 
lightest because of the renormalization group property of the gaugino 
masses, $M_A(\mu_R) = (g_A^2(\mu_R)/g_A^2(M_*))M_A(M_*)$, where $g_A$ 
($A=1,2,3$) are the $U(1)_Y$, $SU(2)_L$ and $SU(3)_C$ gauge couplings. 
This implies that the LSP and NLSP are determined by the competition 
between the right-handed sleptons, bino, and gravitino, which 
may also be lighter than the other superparticles.

The mass ordering between the right-handed sleptons, bino, and 
gravitino depends on the mechanism generating the gaugino masses 
and the universal contributions to the sfermion masses.  Here we 
consider three representative cases.  The first is the simplest 
case that all the operators of Eq.~(\ref{eq:L_soft-M*}) exist 
with all $\eta_A$ and tilde parameters of order unity, except that 
$\tilde{\zeta}$ is somewhat smaller (to suppress dangerous low 
energy processes).  The second is that the theory does not contain 
an elementary singlet at $M_*$ ($> M_U$), so that $\eta_A = 
\tilde{\kappa}_\mu = \tilde{\eta}_H = \tilde{\eta}_\Phi = 
\tilde{\zeta} = 0$, and the gaugino and scalar masses are generated 
by gauge mediation with the messenger scale of order the unification 
scale, $M_U$.  An interesting aspect of this theory is that the 
$\mu$ term can be generated from the interaction ${\cal L} \approx 
\int\! d^4\theta\, (H_u H_d + {\rm h.c.})$ via supergravity 
effects, which are comparable to the gaugino and scalar masses: 
$\mu \approx F_X/M_{\rm Pl} \approx (g_A^2/16\pi^2) F_X/M_U 
\approx m_{\lambda,\tilde{q},\tilde{l}}$, where $M_{\rm Pl} 
\approx 10^{18}~{\rm GeV}$ is the reduced Planck scale, and 
$m_{\lambda,\tilde{q},\tilde{l}}$ represents the gaugino, squark 
and slepton masses ($\tilde{\kappa}_b$ must be suppressed for 
$M_*$ smaller than $M_{\rm Pl}$).  The third is a class of theories 
considered in Ref.~\cite{Ibe:2007km}, where $M_* \approx M_U$, 
and the gaugino and universal scalar masses arise from low energy 
gauge mediation.

The right-handed slepton mass-squared, $m_E^2$, and the bino mass, 
$M_1$, at the weak scale are given in terms of their values, $m_{E,H}^2$ 
and $M_{1,H}$, at some high energy scale $M_H$ by
\begin{eqnarray}
  m_E^2 &\simeq& m_{E,H}^2 + \frac{2}{11} 
    \left( 1 - \frac{g_1^4}{g_{1,H}^4} \right) |M_{1,H}|^2,
\label{eq:mE2-sol-RGE} \\
  M_1 &\simeq& \frac{g_1^2}{g_{1,H}^2} |M_{1,H}|,
\label{eq:M1-sol-RGE}
\end{eqnarray}
where $g_1$ and $g_{1,H}$ are the $U(1)_Y$ gauge couplings at the 
weak scale and $M_H$, respectively.  In the first case described 
above, we take $M_H \approx M_*$ and $m_{E,H}^2 \approx 0$ for 
$\tilde{e}_R$ and $\tilde{\mu}_R$.  Neglecting model-dependent 
effects above $M_U$, we can set $M_H \approx M_U$, and we find 
that $m_E^2 < M_1^2$ at the weak scale for these particles, i.e. 
$\tilde{e}_R$ and $\tilde{\mu}_R$ are lighter than $\tilde{B}$. 
The mass of $\tilde{\tau}_R$ depends on the sign and size of 
$m_{E,H}^2 \approx \epsilon_{E_3}^2 |M_{\rm SUSY}|^2$, and may 
be lighter or heavier than $\tilde{e}_R$, $\tilde{\mu}_R$.  In 
the case where the origin of the gaugino and sfermion masses is 
gauge mediation, as in the second and third cases above, we should 
take $M_H$ to be the messenger scale, $M_{\rm mess}$.  We find 
that $\tilde{B}$ is lighter than $\tilde{e}_R$ and $\tilde{\mu}_R$ 
for $M_{\rm mess} \approx M_U$, but the opposite is possible for 
lower $M_{\rm mess}$, depending on the number of messenger fields. 
The mass of $\tilde{\tau}_R$, again, depends on $m_{E,H}^2$.

The gravitino mass is given by
\begin{equation}
  m_{3/2} \simeq \frac{F_X}{\sqrt{3}M_{\rm Pl}},
\label{eq:m32}
\end{equation}
which should be compared to the gaugino and sfermion masses.  In 
the case that all the operators of Eq.~(\ref{eq:L_soft-M*}) exist 
(except for the superpotential ones) with order one $\eta_A$ and 
tilde parameters, the gaugino and sfermion masses are given by
\begin{equation}
  m_{\lambda,\tilde{q},\tilde{l}} \approx \frac{F_X}{M_*}.
\label{eq:mSUSY}
\end{equation}
We consider that $M_*$ is at least as large as $M_U$ to preserve 
successful gauge coupling unification and at most of order 
$M_{\rm Pl}$ to stay in the field theory regime with weakly 
coupled gravity. This then leads to
\begin{equation}
  \frac{M_U}{M_{\rm Pl}} m_{\lambda,\tilde{q},\tilde{l}} 
    \simlt m_{3/2} \simlt m_{\lambda,\tilde{q},\tilde{l}},
\label{eq:m32-range}
\end{equation}
where $M_U/M_{\rm Pl} \approx 10^{-2}$.  Note that order one 
coefficients are omitted in Eq.~(\ref{eq:m32-range}), so that the 
gravitino can be heavier than some (or all) of the superparticles 
in the supersymmetric standard model sector.  Nevertheless, 
a natural range for the gravitino mass is below the typical 
superparticle mass by up to two orders of magnitude.

The gravitino mass in the other two cases also falls in the 
range of Eq.~(\ref{eq:m32-range}).  In our second example, the 
superparticle masses are given approximately by $(g_A^2/16\pi^2) 
F_X/M_U \approx F_X/M_{\rm Pl}$, leading to $m_{3/2} \approx 
m_{\lambda,\tilde{q},\tilde{l}}$.  The third example has 
superparticle masses of order $(g_A^2/16\pi^2) F_X/(M_U^2/M_{\rm Pl})
\approx F_X/M_U$, leading to $m_{3/2} \approx (M_U/M_{\rm Pl})\, 
m_{\lambda,\tilde{q},\tilde{l}} \approx 10^{-2} 
m_{\lambda,\tilde{q},\tilde{l}}$.

We conclude that the mass ordering between the right-handed sleptons, 
bino, and gravitino is model dependent.  We find, however, that a natural 
range for the gravitino mass is given by Eq.~(\ref{eq:m32-range}).%
\footnote{The gravitino mass can be outside this range.  A smaller 
 gravitino mass could arise, for example, if the physics of flavor 
 and supersymmetry breaking occurs below $M_U$ consistently with gauge 
 coupling unification.  A larger gravitino mass is also possible if 
 the couplings between $X$ and the matter and Higgs fields are somehow 
 suppressed.  For example, if the $X$ field carries a suppression factor 
 $\epsilon_X$ then the gravitino mass is enhanced by $\epsilon_X^{-1}$.}
Thus, it is plausible that the LSP is the gravitino with mass smaller 
than the typical superparticle mass by a factor of a few to a few 
hundred.

\section{Signatures at the LHC}
\label{sec:LHC}

Signatures of flavorful supersymmetry at the LHC depend strongly on 
the mass ordering between the right-handed sleptons, $\tilde{l}_R 
= \tilde{e}_R, \tilde{\mu}_R, \tilde{\tau}_R$, the bino, $\tilde{B}$, 
and the gravitino, $\tilde{G}$.  Based on signatures at the LHC, 
the six possible orderings can be classified into three cases.
\begin{flushleft}
{\bf (a)}\,\, {\boldmath $m_{\tilde{G}} < m_{\tilde{l}_R} 
  < m_{\tilde{B}}$}{\bf :}
\end{flushleft}
One of the right-handed sleptons is the NLSP, which decays into the 
LSP gravitino.  The lifetime is given by
\begin{equation}
  \tau_{\tilde{l}_R} \simeq 
    \frac{48\pi\, m_{\tilde{G}}^2 M_{\rm Pl}^2}{m_{\tilde{l}_R}^5} 
    \left( 1 - \frac{m_{\tilde{G}}^2}{m_{\tilde{l}_R}^2} \right)^{-4},
\label{eq:tau-NLSP}
\end{equation}
which is longer than $\sim 100~{\rm sec}$ for $m_{3/2}$ in the range 
of Eq.~(\ref{eq:m32-range}).  Signatures are therefore stable charged 
tracks inside the main detectors, as well as the late decay of the 
lightest slepton in a stopper which could be placed outside the 
main detector.
\begin{flushleft}
{\bf (b)}\,\, {\boldmath $m_{\tilde{l}_R} < m_{\tilde{B}},\,  
 m_{\tilde{G}}$}{\bf :}
\end{flushleft}
One of the right-handed sleptons is the LSP, leaving charged tracks 
inside the detector.  This case, however, has the cosmological problem 
of charged stable relics.
\begin{flushleft}
{\bf (c)}\,\, {\boldmath $m_{\tilde{B}},\, m_{\tilde{G}} 
  < m_{\tilde{l}_R}$}\,\, {\bf or}\,\, {\boldmath $m_{\tilde{B}} 
  < m_{\tilde{l}_R} < m_{\tilde{G}}$}{\bf :}
\end{flushleft}
A slepton decays into a bino and a lepton inside the detector, so 
that characteristic signatures are conventional missing energy events. 
Intrinsic flavor violation in the supersymmetry breaking masses, 
however, may still be measured by looking at various distributions 
of kinematic variables.

\subsection{Long-lived slepton}
\label{subsec:long-lived}

We begin our discussion with case~(a) above, in which (one of) the 
right-handed sleptons is the NLSP decaying into the LSP gravitino. 
The lifetime of the decay, however, is longer than $\sim 100~{\rm sec}$, 
so that the NLSP is stable for collider analyses.

In the LHC, a stable charged particle interacts with the detector in 
much the same way as a muon.  Therefore its momentum can be measured 
in both the inner tracker and the muon system.  Because of its large 
mass, however, it will generally move slower than a muon.  If its 
speed is in the range $0.6 \simlt \beta \simlt 0.8$, then its mass 
will be measured to a precision of order a few percent~\cite{ATLAS-TDR,%
De Roeck:2005bw}.  While not all NLSP's produced have velocity in 
this range, it is reasonable to expect that a substantial fraction 
will.  Even though they are produced from decays of much heavier 
strongly interacting superparticles, there will usually be several 
decay branches, each of which will divide the energy of the event. 
This reasoning is confirmed by more detailed study~\cite{Hamaguchi:2006vu}. 
With a measurement of the NLSP mass, we can do full reconstruction 
of decay chains which will reduce the uncertainty in the NLSP mass 
to of order $0.1\%$~\cite{Hinchliffe:1998ys},%
\footnote{This precision can be achieved if the systematic 
 uncertainties are $\sim 100~{\rm MeV}$ and the squarks and 
 gluinos are not too heavy.} 
and can measure more parameters of the low energy theory.

To determine the relationship between supersymmetry breaking and 
flavor physics, a critical measurement is the flavor content of 
the leptonic NLSP.  Because flavor mixing is generically suppressed 
by $\epsilon$ factors, the NLSP will be mostly of a single 
flavor. In addition, the NLSP is right handed, so the coupling 
to the charginos will be small, except possibly the Higgsino to 
$\tilde{\tau}_R$.  The coupling to the neutralino with mostly 
$\tilde{B}$ content, however, will be large, so the NLSP will 
usually be produced with a charged lepton of the same flavor. 
Therefore, we can look for events with only two isolated leptons 
and two NLSP's, and a (large) number of jets.  Most such events 
will have leptons of the same flavor as the NLSP.  The high 
effective mass of the event should significantly reduce the 
standard model backgrounds (mostly coming from fakes in events 
with heavy flavors plus jets or electroweak gauge bosons plus 
jets).  Further background rejection, including supersymmetric 
and combinatoric, will be possible by reconstructing the masses 
of the intermediate particles.  This could be complicated if 
the NLSP is mostly $\tilde{\tau}_R$, because we cannot fully 
reconstruct $\tau$'s, but the invariant mass can still be 
reconstructed and the flavor of the NLSP can be identified.

We now analyze the possibility of probing the flavor properties 
of the heavier sleptons.  In particular, we focus on the situation 
where $\tilde{l}_1$ and $\tilde{l}_2$ are mostly $\tilde{e}_R $ and 
$\tilde{\mu}_R$.  We consider, for definitiveness, the case where 
$\tilde{e}_R$ is the NLSP, although the same analysis applies if 
$\tilde{\mu}_R$ is the NLSP.  As shown in section~\ref{subsec:ordering}, 
it is likely that $\tilde{\mu}_R$ is only $\epsilon_{E_2}^2 M_{\rm SUSY} 
\approx $ a few GeV heavier than the NLSP, so the decay of $\tilde{B}$ 
will produce $\tilde{e}_R$ about half the time and $\tilde{\mu}_R$ 
just as often.  When a $\tilde{\mu}_R$ is produced, it will decay 
into a $\tilde{e}_R$ and two leptons.  The leptons produced in this 
decay will be soft in the $\tilde{\mu}_{R}$ rest frame, having energy 
of order only a few GeV, but in general the system will be boosted. 
This possibly poses a problem: for a fast $\tilde{\mu}_R$ the 
leptons will be harder but highly collimated with the NLSP track, 
$\theta \simlt 0.1$, while for a slow $\tilde{\mu}_R$ the opening 
angle will be larger but the leptons will have low $p_{T}$. One 
expects that in some intermediate kinematic regime a reconstruction 
may be feasible, but a detailed study of this issue is beyond the 
scope of this paper.  If this reconstruction turns out to be possible, 
one can look for events where one $\tilde{\mu}_R$ is produced, 
decaying to $\tilde{e}_R$.  These events will have two hard leptons, 
two soft leptons, and two NLSP's.  This event topology should make 
it possible to measure the mass difference between the two lightest 
sleptons, as well as to provide information on the flavor content 
of the (N)NLSP by looking at the flavor of the four leptons.

In the region of parameter space where $\alpha_\beta/\alpha_l \ll 1$, 
the flavor non-universal contribution will be very small and the 
sleptons will be degenerate.  In this co-NLSP region the decay of 
one slepton into another is suppressed because the decay into charged 
sleptons is not kinematically allowed and the right-handed sleptons 
do not couple to neutrinos.  In this region, all three right-handed 
sleptons are long lived, and extracting information on intrinsic 
flavor violation in the supersymmetry breaking parameters requires 
careful analyses.  Since this is a small region of parameter space, 
we do not focus on it here.

The above analysis was for case~(a) where the slepton was the NLSP and 
the gravitino the LSP, but it also applies to case~(b) where the slepton 
is the LSP.  While this scenario is disfavored cosmologically by limits 
on charged relics, the situation could be ameliorated by, for example, 
slight $R$ parity violation in the lepton sector, along with a solution 
to the dark matter problem independent of supersymmetry.%
\footnote{An alternative possibility is that the slepton decays into 
 the axino, the fermionic superpartner of the axion, with a lifetime 
 (much) longer than the collider time scale.  The phenomenology of this 
 scenario is similar to the case with a gravitino LSP.}

\subsection{Late decay of the long-lived slepton}
\label{subsec:late-decay}

In order to determine the lifetime of the NLSP slepton, we would like 
to observe its decays.  The NLSP's produced with $\beta \simlt 0.4$ 
will be stopped within the detector.  One can then detect NLSP decays 
by looking for particles which do not point back to the interaction area. 
Another possibility is that the NLSP will be stopped in the rock just 
outside the detector, and then some of the decay products will re-enter 
the detector.  Unfortunately, very few NLSP's will be produced with 
low enough velocity, and one has to deal with cosmic neutrino background. 
A further possibility is to use the tracker to determine where 
in the surrounding rock the NLSP is stopped.  If the lifetime is 
longer than a few weeks, we could then extract pieces of the rock 
where the NLSP is stopped and study the decay in a more quiet 
environment~\cite{De Roeck:2005bw}.  This will also not have very 
many events, but it will allow very precise measurement of the 
mass and decay properties of the NLSP.

In addition, a large stopper detector can be built outside 
the main detector to trap the NLSP's and measure their decay 
products~\cite{Hamaguchi:2006vu,Hamaguchi:2004df}.  Conventional 
scenarios only consider a $\tilde{\tau}_R$ NLSP, but in flavorful 
supersymmetry the NLSP could be one of the other sleptons, which 
would decay to a monochromatic electron or muon.  This would 
make it very easy to (i) measure the mass of the gravitino given 
the mass of the NLSP measured in the collider, (ii) measure 
the lifetime of the NLSP by counting the number of decays as 
a function of time, and (iii) test supergravity relations such 
as Eq.~(\ref{eq:tau-NLSP})~\cite{Buchmuller:2004rq}, and make 
sure that the gravitino is indeed the LSP.  The stopper detector 
proposed in Ref.~\cite{Hamaguchi:2006vu} did not include a magnetic 
field, so it could not measure the energy of muons, only of 
electrons and taus.  Perhaps this design can be modified to 
include a magnetic field to measure the momentum of the muons.

A stopper detector can very precisely measure the flavor content of 
the NLSP.  If a sufficient number of NLSP's are trapped and there 
is flavor mixing, then a few of the NLSP's will decay to a lepton 
with different flavor.  This occurs in a very clean environment so 
there should be almost no fakes once the accelerator is turned off. 
A stopper detector can very efficiently separate electrons from muons, 
and it can use the monochromatic spectrum of the first two generation 
slepton decays to distinguish $\tau$ decay products.  Mixing angles 
as small as about $10^{-2}$ can be measured~\cite{Hamaguchi:2004ne}. 
The main background comes from cosmic neutrino events, but those 
should all have much lower energy than the NLSP decays.

\subsection{Neutralino (N)LSP}
\label{subsec:chi-NLSP}

Finally we consider case~(c) where the neutralino is lighter 
than the sleptons.  With this spectrum, all sleptons will decay 
promptly, and measuring flavor violation is more difficult. 
Because the neutralino will escape the detector without interacting, 
every event has missing energy, making event reconstruction much 
more difficult.  For direct slepton production one is forced to 
use kinematic variables such as $M_{T2}$~\cite{Lester:1999tx}, 
but they require very high statistics.  The low Drell-Yan production 
cross section quickly prevents this strategy as the slepton mass 
is increased.  The $\tilde{\tau}_R$ is expected to be very split 
from the other two generations, but looking for $\tau$'s means 
even more particles contributing to missing energy.

We are then driven to study lepton flavor violation in cascade decays 
by looking at multiple edges in flavor-tagged dilepton invariant mass 
distributions, along the lines of Refs.~\cite{Bartl:2005yy,CMS-TDR}. 
This method requires sizable flavor violating couplings and will 
probe both those and any modifications to the slepton spectrum. 
However, in order to perform this study with right-handed sleptons, 
they must be produced by the second lightest neutralino, $\chi_2^0$, 
which will be mostly wino, so it has a small branching fraction 
to right-handed sleptons, typically of order $1\%$.  On the other 
hand, the $\chi_2^0$ and the left-handed sleptons are expected to 
be in the same mass range.  So if the spectrum is such that the 
left-handed sleptons are lighter than $\chi_2^0$, then the branching 
ratio of $\chi_2^0$ to $\tilde{l}_L$ will be large.  One can then 
repeat the analysis of section~\ref{subsec:ordering} in the left-handed 
slepton sector and use the methods of Ref.~\cite{Bartl:2005yy} 
to probe flavor violation.

\section{Conclusions}
\label{sec:concl}

Weak scale supersymmetry provides elegant solutions to many of the 
problems of the standard model, but it also generically gives rise 
to excessive flavor and $CP$ violation.  While most existing models 
assume that the mechanism of mediating supersymmetry breaking to the 
supersymmetric standard model sector is flavor universal, we have 
shown that this is not necessary to satisfy all low energy flavor 
and $CP$ constraints.  We have considered a scenario, flavorful 
supersymmetry, in which the mechanism that suppresses the Yukawa 
couplings also suppresses flavor changing interactions in the 
supersymmetry breaking parameters.  We find that a broad region 
of parameter space is allowed, as long as the superpotential 
couplings generating scalar trilinear interactions are suppressed 
or the superparticles have masses of at least a TeV.

The flavorful supersymmetry framework can lead to mass splitting 
among different generations of squarks and sleptons much larger 
than in conventional scenarios.  This has interesting implications 
on collider physics.  In particular, the mass ordering and splitting 
among the three right-handed sleptons, which are expected to be the 
lightest sfermion species, can easily differ significantly from the 
conventional scenarios.  Signatures at colliders depend strongly 
on the species of the LSP and the NLSP, and we have argued that 
these are likely to be one of the right-handed sleptons, the bino, 
or the gravitino.  The gravitino mass is typically in the range 
$10^{-2} m_{\lambda,\tilde{q},\tilde{l}} \simlt m_{3/2} \simlt 
m_{\lambda,\tilde{q},\tilde{l}}$, where $m_{\lambda,\tilde{q},\tilde{l}}$ 
is a characteristic superparticle mass, so that it is plausible to 
expect that the LSP is the gravitino with mass smaller than the 
typical superparticle mass by a factor of a few to a few hundred.

In the case that the lightest right-handed slepton is lighter than 
the bino, we expect to see the dramatic presence of long-lived charged 
particles at the LHC.  This allows us to do full reconstruction of 
decay chains and reduce the uncertainty in the NLSP mass determination. 
Moreover, it is natural to expect that the lightest right-handed slepton 
is, in fact, the NLSP decaying into the gravitino with the lifetime 
longer than $\sim 100~{\rm sec}$.  Because of intrinsic flavor 
violation in the superparticle masses in flavorful supersymmetry, 
the NLSP is not necessarily $\tilde{\tau}_R$ but can be $\tilde{e}_R$ 
or $\tilde{\mu}_R$, leading to the spectacular signature of monochromatic 
electrons or muons in a stopper detector.  This provides a simple 
method to measure the gravitino mass, as well as the lifetime and 
the flavor content of the NLSP, and will be a smoking gun signature 
for flavorful supersymmetry.  In general, flavorful supersymmetry 
predicts flavor violation in both production and decay of sleptons. 
Precision measurements of these processes will also test the flavor 
content of the sleptons.  While these precision measurements are 
difficult at the LHC, they can be done by using certain event 
topologies, regardless of the LSP species.  Further study of flavor 
violation can also be done at a future linear collider.

The origin of the flavor structure is a deep mystery both in the 
context of the standard model and beyond the standard model.  The 
framework of flavorful supersymmetry allows the LHC to probe this 
physics which may lie at a scale close to the Planck scale.  Precise 
study of processes such as the ones discussed in this paper will be 
crucial to uncover the mechanism that leads to the distinct flavor 
pattern we see in nature.

\begin{flushleft}
{\bf Note Added:}

While completing this paper, we received Ref.~\cite{Feng:2007ke} 
which discusses similar ideas.
\end{flushleft}

\section*{Acknowledgments}

This work was supported in part by the U.S. DOE under Contract 
DE-AC02-05CH11231, and in part by the NSF under grant PHY-04-57315. 
The work of Y.N. was also supported by the NSF under grant PHY-0555661, 
by a DOE OJI, and by the Alfred P. Sloan Research Foundation. 
The work of D.S. was supported by the Alcatel-Lucent Foundation.

\end{document}